\documentclass[twocolumn, showpacs,amsmath,amssymb]{revtex4}
\usepackage{graphicx}
\usepackage{dcolumn}
\usepackage{bm}
\usepackage{stackrel}
\usepackage{amsmath}
\usepackage{amsfonts}
\usepackage{amssymb}
\usepackage{amsthm}
\usepackage{epstopdf}
\usepackage{longtable}
\usepackage{array}
\usepackage{pifont}
\usepackage{latexsym}
\usepackage[latin1]{inputenc}
\usepackage{bm}
\usepackage{textcomp}
\usepackage{subfigure}
\usepackage{braket}
\usepackage{wasysym}
\usepackage{ulem}
\usepackage{color, soul}
\newcolumntype{.}{D{.}{.}{-1}}

\begin{document}

\author{T. Zanon-Willette\footnote{E-mail address: thomas.zanon@upmc.fr\\}}

\affiliation{LERMA, Observatoire de Paris, PSL Research University, CNRS, Sorbonne Universités, UPMC Univ. Paris 06, F-75005, Paris, France}
\author{V.I. Yudin and A.V. Taichenachev}
\affiliation{Institute of Laser Physics, Siberian Branch of Russian Academy of Sciences, Novosibirsk 630090, Russia,
Novosibirsk State University, Novosibirsk 630090, Russia,
and Novosibirsk State Technical University, Novosibirsk 630092, Russia}

\date{\today}

\preprint{APS/123-QED}

\title{Generalized Hyper-Ramsey Resonance with separated oscillating fields}

\begin{abstract}
An exact generalization of the Ramsey transition probability is derived to improve ultra-high precision measurement and quantum state engineering when a particle is subjected to independently-tailored separated oscillating fields. The phase-shift accumulated at the end of the interrogation scheme offering high-level control of quantum states throughout various laser parameters conditions. The Generalized Hyper-Ramsey Resonance based on independent manipulation of interaction time, field amplitude, phase and frequency detuning is presented to increase the performance of next generation of atomic, molecular and nuclear clocks, to upgrade high resolution frequency measurement in Penning trap mass spectrometry and for a better control of light induced frequency shifts in matter wave interferometers or quantum information processing.
\end{abstract}


\maketitle

\section{Introduction}

\indent Precision measurement plays a critical role in many fields of physics such as metrology
and fundamental tests of physical theories. Laser pulsed spectroscopy is one technique capable of realizing such
precision measurements. It is now a universal tool to investigate interaction between light and matter in quantum clocks \cite{Rosenband:2008,Wineland:2013,Hinkley:2013,Bloom:2014,LeTargat:2013}, in cavity quantum electrodynamics experiments \cite{Brune:1990,Gleyzes:2007,Haroche:2013} and in atomic, molecular and neutron interferometry \cite{Cronin:2005,Berman:1996,Klepp:2014}.

To improve the resolution of frequency measurements in the atomic and molecular beam resonance method devised by I.I. Rabi \cite{Rabi:1938}, N.F. Ramsey proposed to replace the single oscillatory field by a double microwave excitation pulse separated by a free evolution without any electromagnetic field perturbation \cite{Ramsey:1950,Ramsey:1956}.\\
The low sensitivity of Ramsey spectroscopy to field inhomogeneities inducing light-shifts on the probing atomic transition has drastically impacted the time and frequency metrology \cite{Ramsey:1990,Vanier:1989,Gibble:1992} leading to microwave standards at the relative $10^{-16}$ level of accuracy \cite{Wynands:2005,Bize:2005}.

Ultra-high resolution frequency measurement has been achieved with very long storage time of Doppler and recoil free quantum particles using laser cooling techniques in ion traps \cite{Rosenband:2008,Margolis:2009} and optical lattice clocks \cite{DereviankoKatori:2011,Ludlow:2015}. The level of 10$^{-18}$ relative accuracy, now almost achieved \cite{Nicholson:2015}, requires a very precise control of atomic or molecular interactions to cancel systematic frequency shifts, whether fermionic or bosonic species are used \cite{Campbell:2009,Lisdat:2009}. With the next generation of quantum clocks, stringent tests of general relativity are possible as well as new applications in geophysics and hydrology \cite{Chou:2010}.
Recently, very high precision measurement has become relevant for mass spectrometry \cite{Blaum:2006} where the highest precision was reached using ions in a Penning trap. The use of Ramsey's method of separated oscillating fields has provided a significant reduction in the uncertainty of the cyclotron frequency and thus of the ion mass of interest combined with a faster acquisition rate \cite{Bollen:1992,George:2007,Eibach:2011}.\\
For the next level of progress in very high precision, the recent Hyper-Ramsey scheme \cite{Yudin:2010,Tabatchikova:2013, Tabatchikova:2015} is a promising evolution of Ramsey pulsed spectroscopy.
Composite pulses inspired by NMR techniques \cite{Levitt:1986} are now used to compensate simultaneously for noise decoherence, pulse area fluctuation and residual frequency offset due to the applied laser field itself. The first or the second pulse of the usual Ramsey sequence can be separated in two or more contiguous sections which yield to probing protocols with more degrees of freedom in order to minimize resonance shifts \cite{Vandersypen:2005,Braun:2014}.
It is also reminiscent of the spin echo technique where a sequence of pulses was originally applied to suppress inhomogeneous effects causing a spin relaxation \cite{Hahn:1950,Carr:1958}. The Hyper-Ramsey method was successfully implemented on a single trapped $^{171}$Yb$^+$ ion demonstrating efficient reduction of the light-shift by several orders of magnitude \cite{Huntemann:2012}.

\begin{figure}[t!!]
\centering%
\resizebox{8.0cm}{!}{\includegraphics[angle=-90]{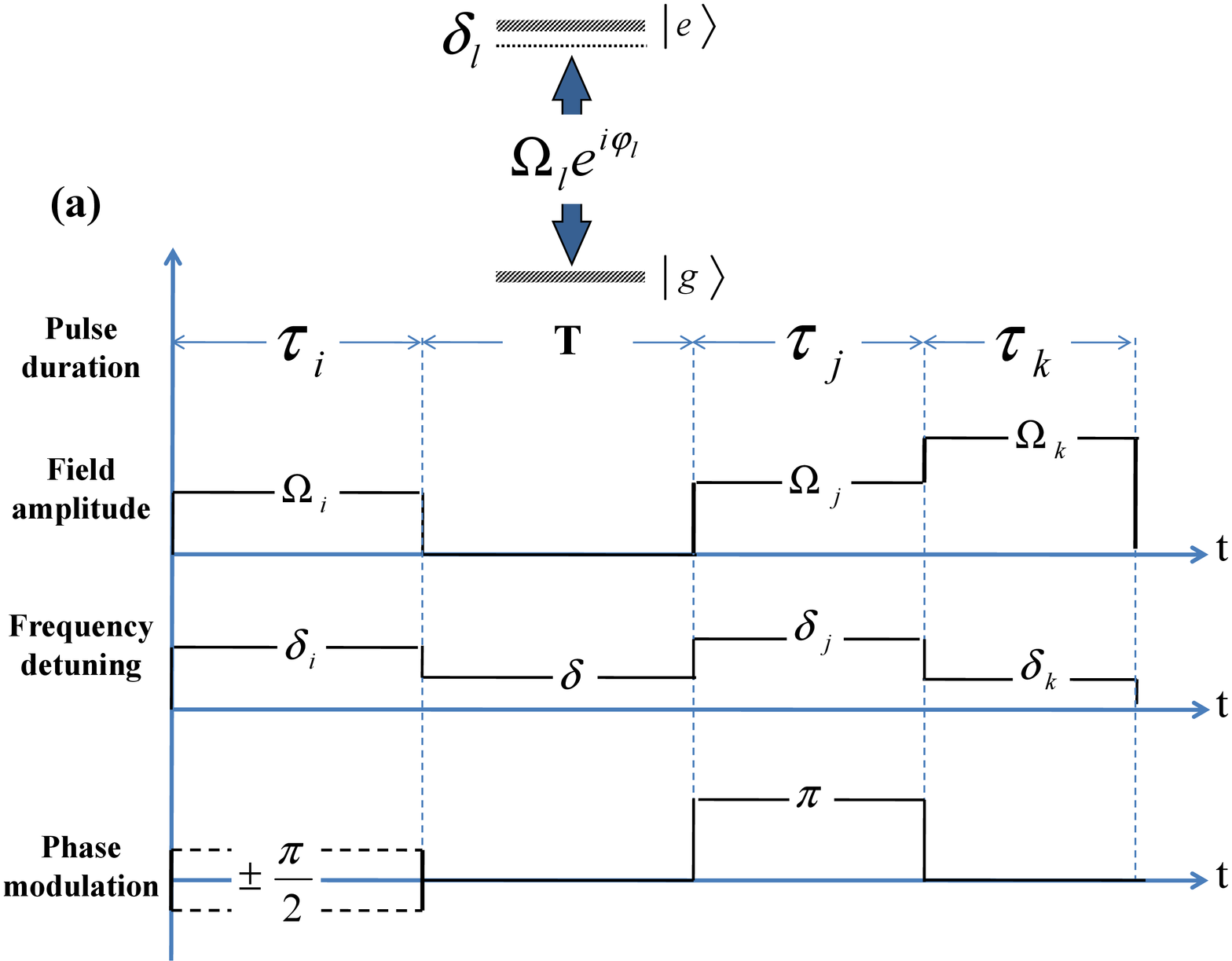}}
\resizebox{8.0cm}{!}{\includegraphics[angle=-90]{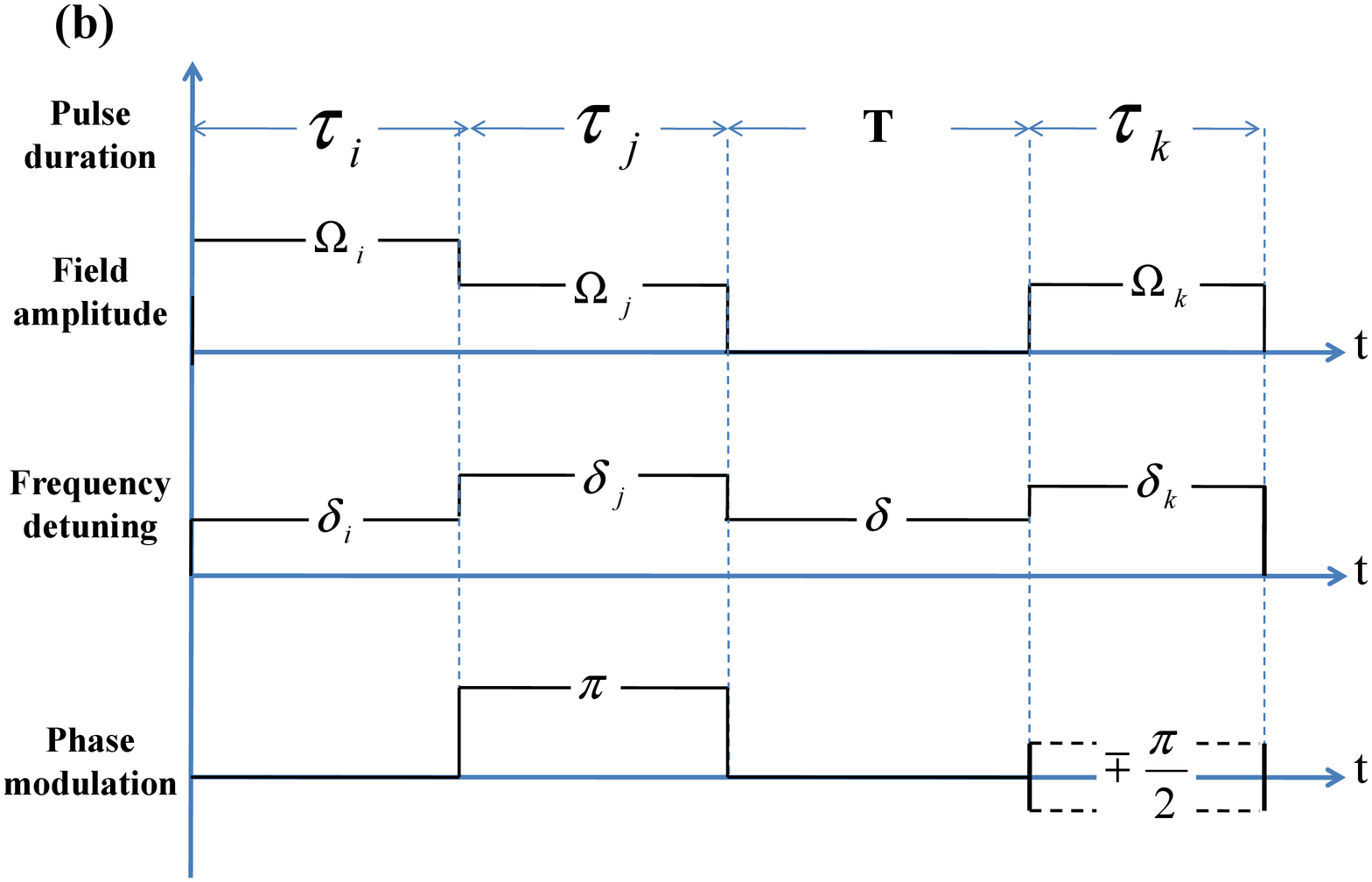}}
\caption{(color online). Quantum system with a narrow $|g\rangle\leftrightarrow|e\rangle$ clock transition probed by a laser pulse excitation. (a) Sequence of composite pulses $\theta_i,\delta T,\theta_j,\theta_k$ and (b) $\theta_i,\theta_j,\delta T,\theta_k$ with specified laser parameters including detuning $\delta_{l}$, complex field amplitude $\Omega_{l}e^{i\varphi_{l}}$, pulse duration $\tau_{l}$  where $l=i,j,k$ and a free evolution time T between pulses. A phase step modulation of the appropriate laser field $\varphi_{l}$ may be applied in both schemes if necessary.}
\label{sequence}
\end{figure}
\begin{figure}[t!!]
\resizebox{8.5cm}{!}{\includegraphics[angle=0]{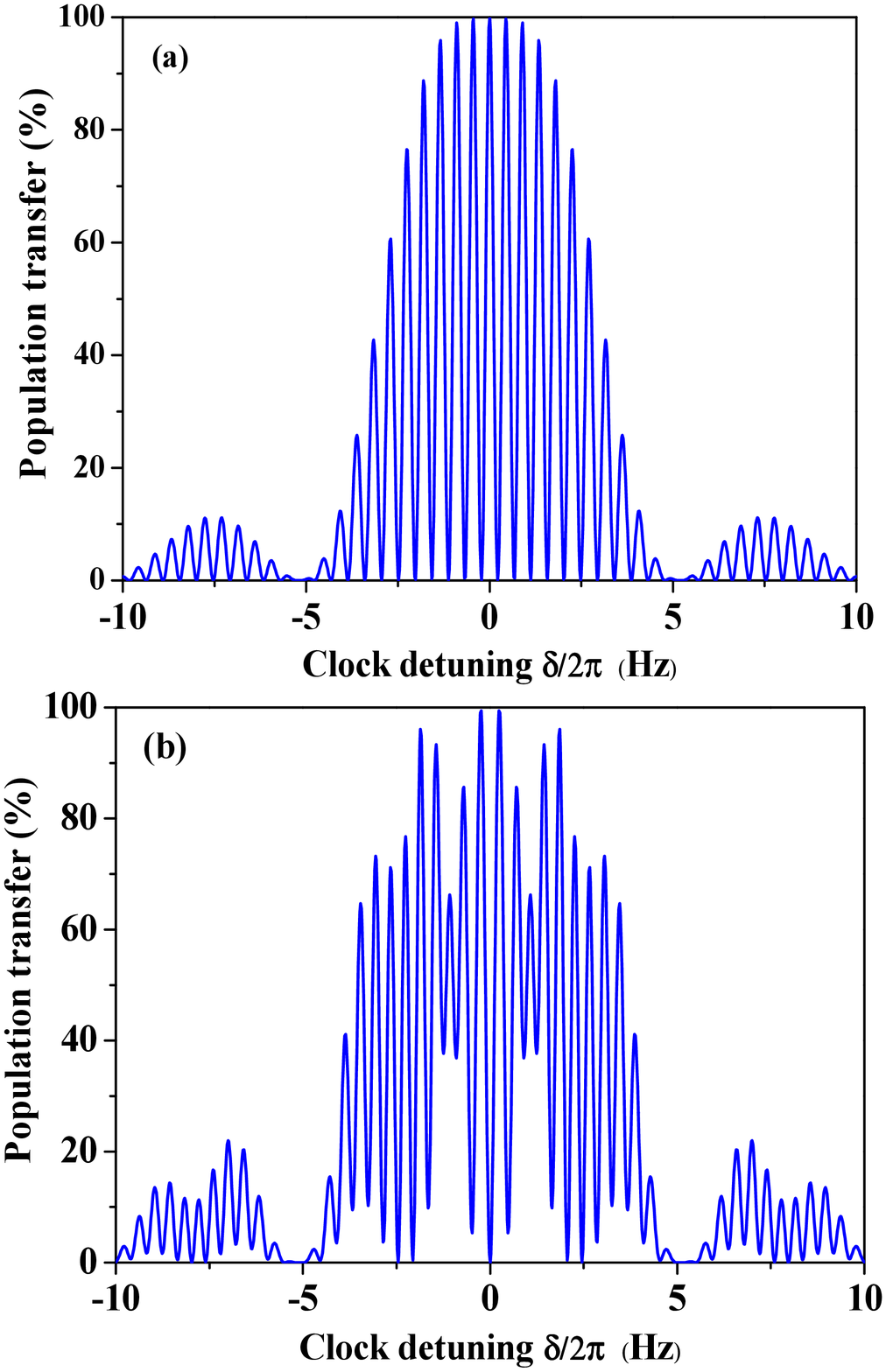}}
\caption{(color online). Generalized Hyper-Ramsey resonances $P_{|g\rangle\mapsto|e\rangle}$ versus the clock detuning $\delta$ computed from Eq.~\ref{Generalized-Hyper-Ramsey-transition} to~\ref{Generalized-Hyper-Ramsey-phase}. (a) Ramsey resonance. (b) Hyper-Ramsey Resonance from all G-H-R protocols reported in Table~\ref{protocol-table}. The optimal Rabi frequency of the laser field is defined as $\Omega=\pi/2\tau$. Pulse duration is $\tau=0.1875$~s with a free evolution time $T=2$~s and $\Delta_{l}=0$ ($l=i,j,k$).}
\label{G-H-R}
\end{figure}

\section{Generalized version of the Ramsey transition probability}

\indent We have established an accurate generalization of the Ramsey interrogation for transition probability. This formalism provides a practical guide to the design, implementation, and interpretation of pulse sequences, and it is thus of considerable importance for high precision spectroscopy widely used in fundamental and applied physics.
The resonance can be coherently excited with weakly allowed transitions \cite{Ludlow:2015}, stimulated two-photon Raman transitions \cite{Zanon:2014,Kasevich:1991,Moler:1992}, by magnetically induced transitions \cite{Taichenachev:2006} or by quadrupolar radio-frequency field interaction in mass spectrometry \cite{Kretzschmar:2007}. It enables the accurate control of energy levels at the end of the interrogation sequence.

\subsection{Analytical form of the transition probability}

Adopting a multi-zone interaction wave-function model \cite{Zanon:2014,Rabi:1945,Varshalovich:1988}, we have derived the analytical form of the generalized transition probability and the phase-shift driving the resonance frequency position around the extremum of the central fringe. This derivation extends the initial work in ref \cite{Yudin:2010}. The generalized transition probability is derived for independent particles interacting with separated and controllable oscillating fields. Field amplitudes, frequency detunings and pulse durations can be manipulated individually within two different probing schemes shown in Fig.~\ref{sequence}(a) and (b).
The sequence of composite pulses described by the analytical expression of the transition probability is investigated to improve control of external perturbations such as light-shifts or residual magnetic fields on the lineshape \cite{Shirley:1963,Greene:1978}. It can be implemented in trapped atom clocks and extended to ion trap mass spectrometers and molecular beams or fountain devices where moving particles are interacting with multiple oscillating fields during a free fly.

\indent The generalized transition probability must to be dependent on pulse area $\theta_{l}=\omega_{l}\tau_{l}/2$ ($l=i,j,k$)
with different driving Rabi amplitudes $\Omega_{l}$ and frequency detunings $\delta_{l}$ via the generalized Rabi frequency $\omega_{l}=\sqrt{\delta_{l}^{2}+\Omega_{l}^{2}}$.
During the pulses, light-shift from off-resonant states may be present and a laser step frequency utilized to rectify the anticipated shift, thus requiring a redefinition of frequency detunings as $\delta_{l}\equiv\delta_{l}\pm\Delta_{l}$. Additional phase inversion of laser fields and nonuniform pulsed excitation conditions modifying the entire spectral lineshape are also included into the computation of the transition \cite{Yudin:2010,Brandin:1994}.
The exact expression of the generalized Ramsey probability for a particle starting from initial state $|g\rangle$ to final state $|e\rangle$ is given in a compact form as:
\begin{equation}
\begin{split}
P_{|g\rangle\mapsto|e\rangle}=&\alpha\left[1+\beta(\Phi)^2\right]\left[1+\frac{2\beta(\Phi)}{1+\beta(\Phi)^2}\cos(\delta T+\Phi)\right],
\label{Generalized-Hyper-Ramsey-transition}
\end{split}
\end{equation}
with a clock frequency detuning $\delta$ during free evolution without light.
We have introduced for convenience the notation
\begin{equation}
\begin{split}
\beta(\Phi)=\beta\sqrt{1+\tan^2\Phi},
\label{reduced-beta}
\end{split}
\end{equation}
where envelopes $\alpha,\beta$ driving the resonance amplitude are respectively defined by
\begin{widetext}
\begin{subequations}
\small{
\begin{align}
\alpha=&\left(1+\frac{\delta_{i}^{2}}{\omega_{i}^{2}}\tan^2\theta_i\right)\left(1+\frac{\delta^{2}_{jk}}{\omega_{jk}^{2}}\tan^{2}\theta_{jk}\right)\left(\frac{\Omega_j}{\omega_j}\tan\theta_j+\frac{\Omega _k}{\omega_k}\tan\theta_k\right)^2\cos^2\theta_i\cos^2\theta_j\cos^2\theta_k\label{alpha},\\
\beta=&\frac{\frac{\Omega_{i}}{\omega_{i}}\tan\theta_i\left(1-\frac{\delta_{j}\delta_{k}+\Omega_{j}\Omega_{k}}{\omega_{j}\omega_{k}}\tan\theta_{j}\tan\theta_{k}\right)}{\left(1-\frac{\delta_{i}\delta_{jk}}{\omega_{i}\omega_{jk}}\tan\theta_i\tan\theta_{jk}\right)\left(\frac{\Omega_j}{\omega_j}\tan\theta_j+\frac{\Omega_k}{\omega_k}\tan\theta_k\right)}\frac{1-\frac{\frac{\delta_{j}}{\omega_{j}}\tan\theta_j+\frac{\delta_{k}}{\omega_{k}}\tan\theta_k}{1-\frac{\delta_{j}\delta_{k}+\Omega_{j}\Omega_{k}}{\omega_{j}\omega_{k}}\tan\theta_{j}\tan\theta_{k}}
\frac{\frac{\delta_{i}}{\omega_{i}}\tan\theta_i+\frac{\delta_{jk}}{\omega_{jk}}\tan\theta_{jk}}{1-\frac{\delta_{i}\delta_{jk}}{\omega_{i}\omega_{jk}}\tan\theta_{i}\tan\theta_{jk}}}
{1+\left(\frac{\frac{\delta_{i}}{\omega_{i}}\tan\theta_i+\frac{\delta_{jk}}{\omega_{jk}}\tan\theta_{jk}}{1-\frac{\delta_{i}\delta_{jk}}{\omega_{i}\omega_{jk}}\tan\theta_{i}\tan\theta_{jk}}\right)^{2}},
\label{beta}
\end{align}}
\end{subequations}
\end{widetext}
including a reduced variable
\begin{equation}
\begin{split}
\frac{\delta_{jk}}{\omega_{jk}}\tan\theta_{jk}=\frac{\left(\delta_j\Omega_k-\Omega_j\delta_k\right)\tan\theta_{j}\tan\theta_{k}}{\Omega_{j}\omega_{k}\tan\theta_{j}+\Omega_{k}\omega_{j}\tan\theta_{k}}.
\end{split}
\label{reduced-variable}
\end{equation}
The phase-shift accumulated after the entire interrogation scheme is
\begin{equation}
\begin{split}
\tan\Phi=&\frac{\frac{\frac{\delta_{j}}{\omega_{j}}\tan\theta_{j}+\frac{\delta_{k}}{\omega_{k}}\tan\theta_{k}}{1-\frac{\delta_j\delta_k+\Omega_j\Omega_k}{\omega_j\omega _k}\tan\theta_{j}\tan\theta_{k}}+\frac{\frac{\delta_{i}}{\omega_{i}}\tan\theta_{i}+\frac{\delta_{jk}}{\omega_{jk}}\tan\theta_{jk}}{1-\frac{\delta_{i}\delta_{jk}}{\omega_{i}\omega_{jk}}\tan\theta_{i}\tan\theta_{jk}}}{1-\frac{\frac{\delta_{j}}{\omega_{j}}\tan\theta_{j}+\frac{\delta_{k}}{\omega_{k}}\tan\theta_{k}}{1-\frac{\delta_j\delta_k+\Omega_j\Omega_k}{\omega_j\omega _k}\tan\theta_{j}\tan\theta_{k}}\frac{\frac{\delta_{i}}{\omega_{i}}\tan\theta_{i}+\frac{\delta_{jk}}{\omega_{jk}}\tan\theta_{jk}}{1-\frac{\delta_{i}\delta_{jk}}{\omega_{i}\omega_{jk}}\tan\theta_{i}\tan\theta_{jk}}}.
\end{split}
\label{Generalized-Hyper-Ramsey-phase}
\end{equation}
From Eqs.~\ref{Generalized-Hyper-Ramsey-transition} to \ref{Generalized-Hyper-Ramsey-phase}, lineshape, population transfer efficiency and frequency-shift affecting the resonance can be used to accurately evaluate in various experimental laser pulse conditions including Rabi excitation, Ramsey and Hyper-Ramsey schemes \cite{Rabi:1938,Ramsey:1950,Yudin:2010}. Note that a second composite sequence of pulses can be realized as proposed in Fig.~\ref{sequence}(b). All previous expressions are still valid by simply exchanging subscripts $i\rightleftarrows k$ with corresponding laser parameters while reversing sign for pulse durations $\tau_{l}\mapsto-\tau_{l}$ and free evolution time $T\mapsto-T$.
We also confirm the consistency of our analytic results by comparing to a numerical solution in the rotating frame under various pulse conditions \cite{Rabi:1954,Jaynes:1955,Goldstein:1980,Feynman:1957,Glass-Maujean:1991}.

To highlight a natural connection between the well-known method of separated oscillatory fields by N. Ramsey \cite{Ramsey:1950,Ramsey:1956,Ramsey:1990} and the generalized formalism presented in this work, we report in the Appendix a straightforward derivation of the standard Ramsey transition probability based on our generalized pulse sequence.

\subsection{Generalized expression of the phase-shift}

The phase-shift given by Eq.~\ref{Generalized-Hyper-Ramsey-phase} is the primary result needed for the precise control of quantum states to suppress any frequency-shift induced by the laser excitation itself. By tailoring phase-shift parameters and engineering the resonance amplitude, it is possible to generate quantum spectroscopy signals for increased resolution or better stability.
Following \cite{Abramowitz:1968}, Eq.~\ref{Generalized-Hyper-Ramsey-phase} can be written into a closed form solution as:
\begin{equation}
\begin{split}
\Phi=&\arctan\left[\frac{\frac{\delta_{j}}{\omega_{j}}\tan\theta_{j}+\frac{\delta_{k}}{\omega_{k}}\tan\theta_{k}}{1-\left(\frac{\delta_j\delta_k+\Omega_j\Omega_k}{\omega_j\omega _k}\right)\tan\theta_{j}\tan\theta_{k}}\right]\\
&+\arctan\left[\frac{\delta_{jk}}{\omega_{jk}}\tan\theta_{jk}\right]+\arctan\left[\frac{\delta_{i}}{\omega_{i}}\tan\theta_{i}\right].
\end{split}
\label{Generalized-Hyper-phase}
\end{equation}
It is thus feasible to eliminate the frequency shift of the central fringe by engineering Eq.~\ref{Generalized-Hyper-phase} with special choices of laser step frequency, pulse duration, and phase inversion to achieve a desired interference minimum.
The key point is to establish some efficient quantum control protocols which compensate for frequency-shift and are robust to small change in pulse area while achieving a highly contrasted population transfer between the targeted states \cite{Braun:2014}. Such quantum engineering of phase-shift has been recently proposed in a Generalized Hyper Raman-Ramsey spectroscopy of a stimulated two-photon forbidden clock transition of strontium $^{88}Sr$ and ytterbium $^{174}Yb$ eliminating the detrimental light-shift contribution \cite{Zanon:2014}.\\
\begin{table}[b!!]
\centering%
\caption{Selected Ramsey (R) and Generalized Hyper-Ramsey (G-H-R) pulses protocols shown in Fig.~\ref{sequence}(a) and (b) where $\Delta_{l}=0$ ($l=i,j,k$). The Rabi frequency of the light field is defined as $\Omega=\pi/2\tau$ and possible phase inversion of the light field during a pulse is indicated. The phase step modulation of the laser field is off.}
\label{protocol-table}
\begin{tabular}{||c|cccc||}
\hline
protocols & parameters & $\theta_i$ & $\theta_j$ & $\theta_k$  \\
\hline
& $\tau_{l}$ & $\tau$ & $0$ & $\tau$  \\
\text{R}  & $\Omega_{l}$  & $\Omega$ & $0$ & $\Omega$  \\
& $\delta_{l}$  & $\delta$ & $0$ & $\delta$   \\
\hline
        & $\tau_{l}$ & $\tau$ & $2\tau$ & $\tau$   \\
        & $\Omega_{l}$ & $\pm\Omega$ & $\mp\Omega$ & $\pm\Omega$   \\
\text{G-H-R} & $\delta_{l}$  & $\delta$ & $\delta$ & $\delta$   \\
          \cline{2-5}
        & $\tau_{l}$ & $\tau$ & $\tau$ & $\tau$  \\
        & $\Omega_{l}$  & $\pm\Omega$ & $\mp2\Omega$ & $\pm\Omega$  \\
        & $\delta_{l}$  & $\delta$ & $2\delta$ & $\delta$  \\
\hline
\end{tabular}
\end{table}
The Generalized Hyper-Ramsey transition probability has been computed for differing pulse protocols reported in Table~\ref{protocol-table}. The standard sequence (a Ramsey protocol (R)) is compared with two others sequences based on nonstandard G-H-R protocols.
The panels of Fig.~\ref{G-H-R} display resonance fringes corresponding to selected protocols in Table~\ref{protocol-table}.
In Fig.~\ref{G-H-R}(a), Ramsey fringes have been simulated using parameters following the R protocol with two $\theta_i=\theta_k=\pi/2$ pulse areas. Hyper-Ramsey fringes have also been simulated with $\theta_i=\pi/2$ and $\theta_j=\pi$, $\theta_k=\pi/2$ pulse areas according to G-H-R protocols leading to the same resonance line shown in Fig.~\ref{G-H-R}(b).

\begin{figure}[t!!]
\resizebox{9.0cm}{!}{\includegraphics[angle=0]{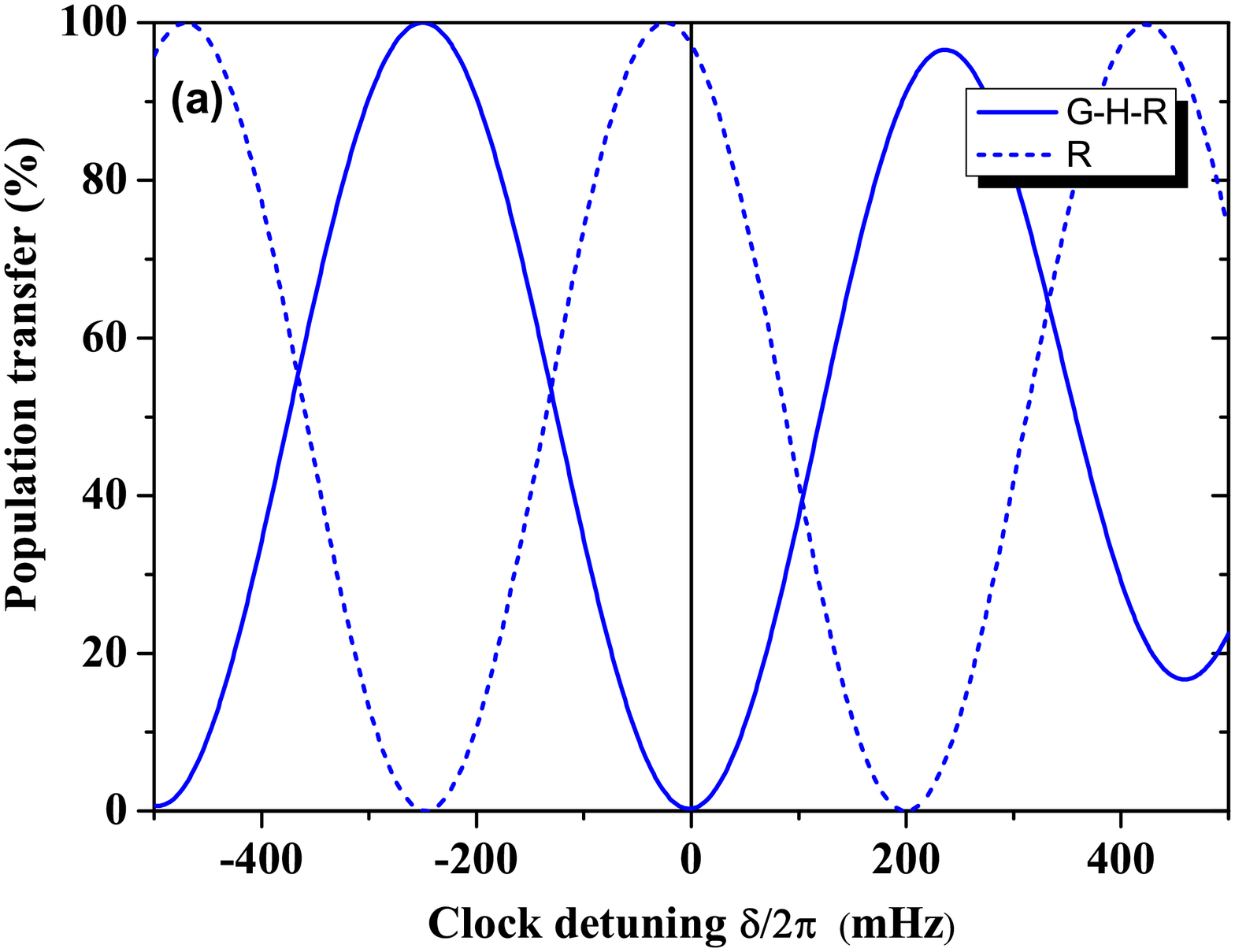}}
\resizebox{9.0cm}{!}{\includegraphics[angle=-90]{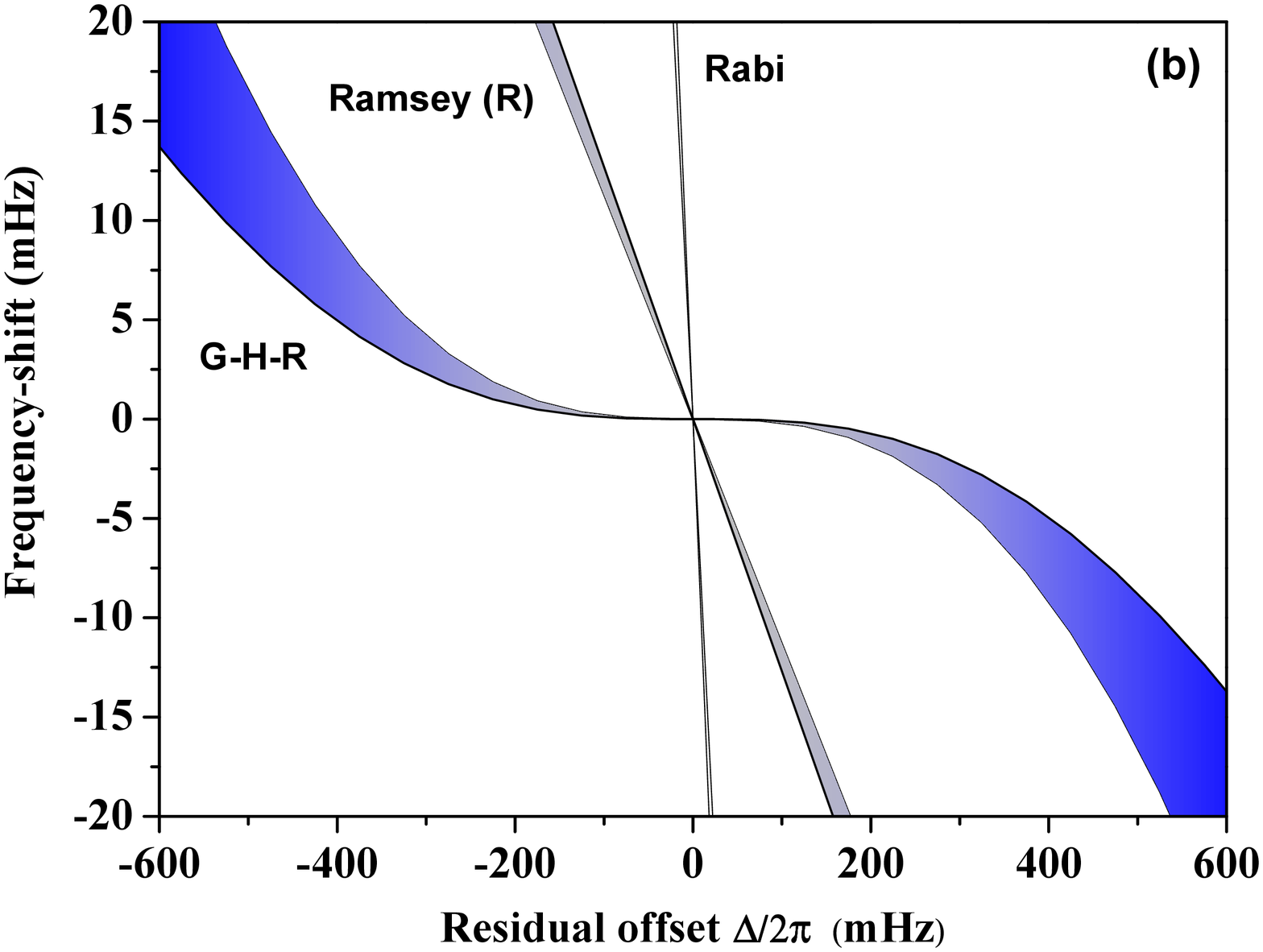}}
\caption{(color online). (a) Small offset fixed around $\Delta/2\pi\sim220$~mHz inducing a frequency shift on the central Ramsey resonance (short dashed line) and the Generalized Hyper-Ramsey resonance (solid line). (b) Central fringe frequency-shift $-\Phi/2\pi T$ based on Eq.~\ref{Generalized-Hyper-Ramsey-phase} (Eq.~\ref{Generalized-Hyper-phase}) as a function of the residual offset $\delta\equiv\Delta$ for all protocols reported in Table~\ref{protocol-table}. All curves are also shown with a $\pm10\%$ pulse area variation (shadow areas between solid lines). Others parameters are identical to Fig.~\ref{G-H-R}.}
\label{G-H-R-shift}
\end{figure}
\begin{figure}[t!!]
\resizebox{9.0cm}{!}{\includegraphics[angle=0]{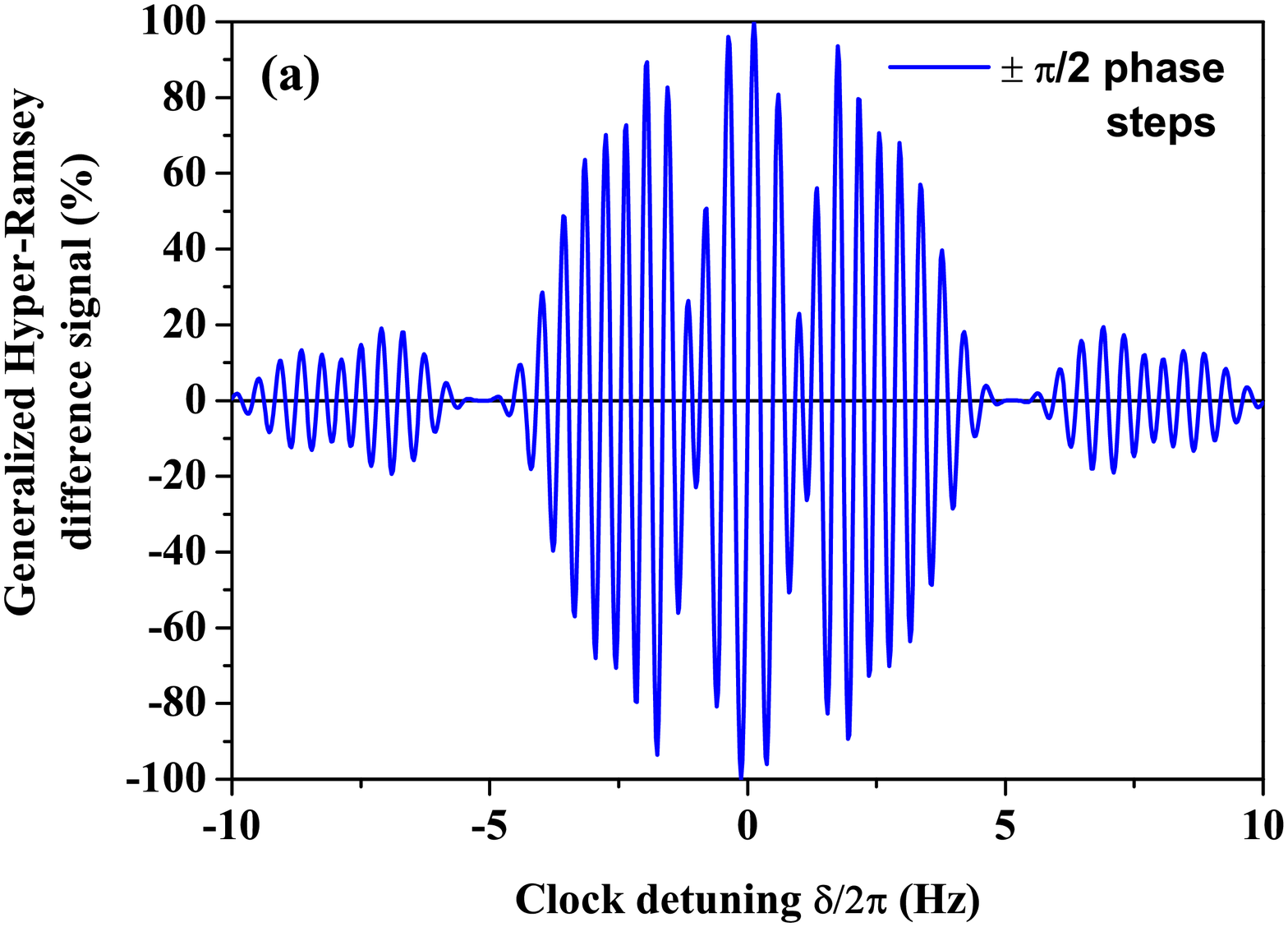}}
\resizebox{9.0cm}{!}{\includegraphics[angle=0]{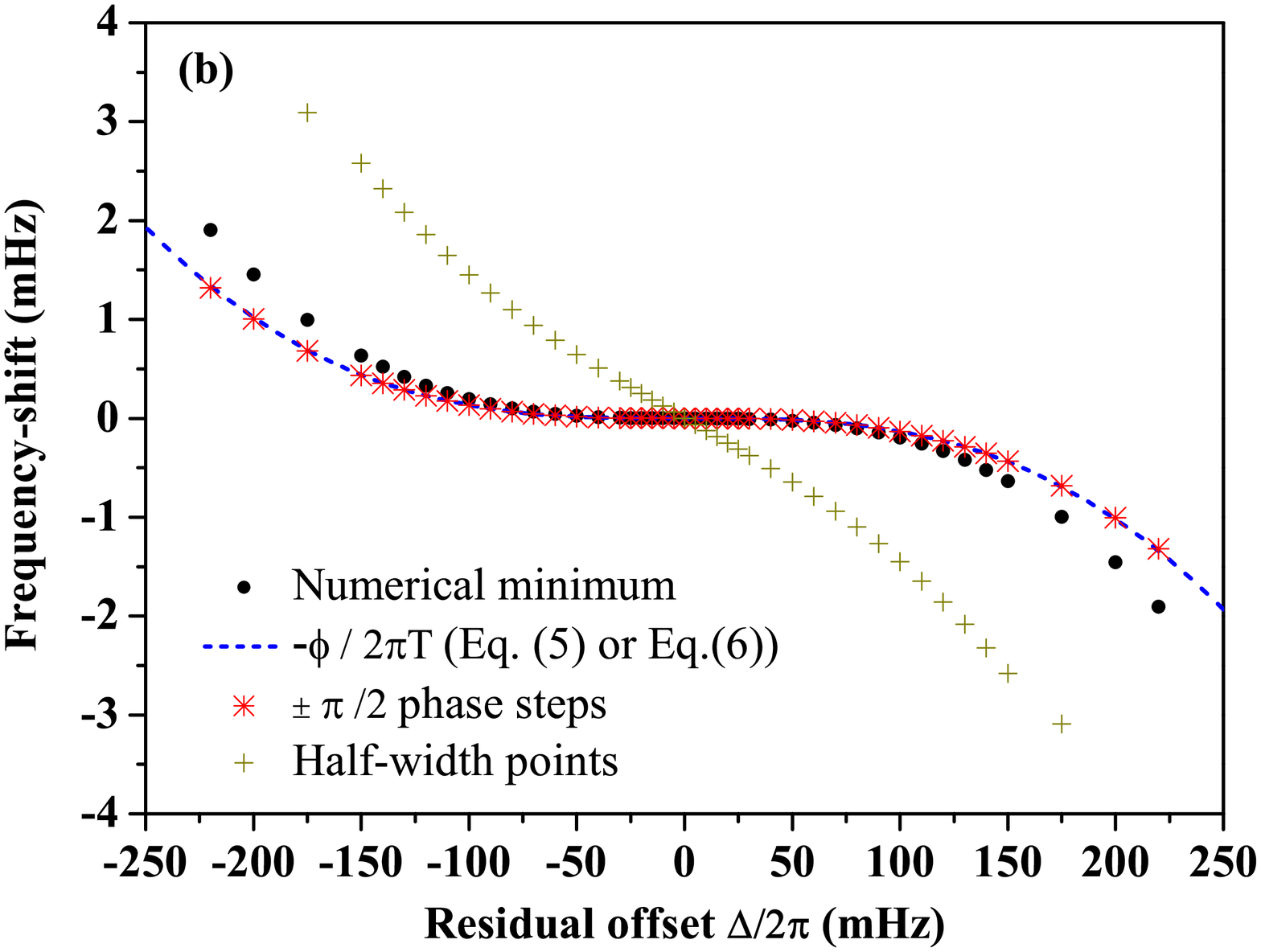}}
\caption{(color online). (a) Dispersive error signal generating by a $\pm\pi/2$ phase step modulation of appropriate laser fields. (b) Evaluation of the frequency-shift versus a residual frequency offset by a numerical tracking of the central fringe extremum ($\bullet$ solid black dots), by the phase step process of Eq.~\ref{Generalized-Hyper-Ramsey-transition} ($\star$ solid stars), by an evaluation based on Eq.~\ref{Generalized-Hyper-Ramsey-phase} (Eq.~\ref{Generalized-Hyper-phase}) with $-\Phi/2\pi T$ (dashed line) or by probing the central fringe at half-width points (solid crosses).}
\label{shift-comparison}
\end{figure}

Non-linear behaviors of the central fringe frequency-shift versus a small frequency perturbation in the clock detuning have been investigated and presented in Fig.~\ref{G-H-R-shift}(a) and \ref{G-H-R-shift}(b). The curves are compared to the linear frequency shift resulting from a single Rabi pulse excitation.
Fig.~\ref{G-H-R-shift}(a) shows the effect of a small external frequency offset on the central fringe position.
In contrast to the normal Ramsey spectroscopy, the Generalized Hyper-Ramsey resonance method provides a vastly reduced sensitivity of the central fringe's frequency shift to external perturbations. As evidenced in Fig.~\ref{G-H-R-shift}(b), small perturbations such as the residual light-shift or magnetic field fluctuations may manifest themselves as laser frequency steps during the spectroscopy pulse sequence, but the central fringe position has a very flat slope for its frequency shift dependence, which is more than one order of magnitude smaller than that for Ramsey.
The combination of $\theta_i=\pi/2$ and $\theta_j=\pi$, $\theta_k=\pi/2$ pulse areas in Eq.~\ref{Generalized-Hyper-phase} compensates all terms over a large residual error in frequency as well as their first and second order sensitivity to frequency fluctuation.
The suppression of the central fringe shift may be finally made insensitive to small pulse area variation by inserting an intermediate "echo" pulse with a sign inversion of the light field only during the intermediate $\theta_j=-\pi$ pulse or during the initial and final $\theta_i=\theta_k=-\pi/2$ pulses (see Table~\ref{protocol-table}). Simultaneously, a laser frequency step may be ultimately applied during pulses to compensate any frequency offset larger than the width of the resonance \cite{Huntemann:2012}.

\subsection{Phase step modulation of the resonance}

\indent The top plot in Fig.~\ref{G-H-R-shift}(a) raises a serious concern for precision metrology. While the minimum of the Ramsey fringe occurs at line center, the first maxima of the G-H-R fringes on either side differ in amplitude by $3-4~\%$. Many precision experiments, among them clock experiments, actually lock to the central feature to stabilize the frequency on the line center. Since locking requires an odd symmetry feature, the exciting laser usually has its frequency modulated to generate the needed signal. However, modulation of an asymmetric resonance would lead to off-center locking, exactly what we are trying to avoid with the G-H-R technique. The amount of shift depends on the modulation details, but a typical modulation usually at half-width of the central fringe is required. If the 50$\%$ population levels are not equally spaced from line center, the resulting modulated line will be slightly pulled out from line center.\\
In order to eliminate the asymmetry effect on the true position of the central fringe, we propose to apply a $\pm\pi/2$  phase step modulation of the cosine function in Eq.~\ref{Generalized-Hyper-Ramsey-transition} as in refs \cite{Ramsey:1951,Letchumanan:2004}. The effect of the laser phase modulation on the transition probability is discussed in the Appendix with a complex wave-function model.
Fig.~\ref{shift-comparison}(a) shows the resulting dispersive error signal generated in this way.
Fig.~\ref{shift-comparison}(b) presents a comparison between different frequency-locking techniques to probe the extremum of the central fringe. Solid black dots report the exact position of the extremum by a numerical tracking of Eq.~\ref{Generalized-Hyper-Ramsey-transition}. By applying a frequency modulation technique and probing the interferences at the half-width points to lock to the center of the resonance, the slight asymmetry of the lineshape reintroduces a weak linear dependance of the frequency-shift with a residual frequency offset. The phase step modulation of Eq.~\ref{Generalized-Hyper-Ramsey-transition} is able to reproduce very well the frequency-shift based on Eq.~\ref{Generalized-Hyper-Ramsey-phase} (Eq.~\ref{Generalized-Hyper-phase}) even in presence of a weak lineshape asymmetry. This phase modulation technique directly produces an error signal with enhanced immunity to residual offset fluctuations \cite{Huntemann:2012,Morinaga:1989,Letchumanan:2004}.

\section{Conclusion}

\indent Although the initial Ramsey's method of separated oscillating fields has proven to be very useful in fundamental and applied physics based on laser pulsed spectroscopy, it still has some fundamental limitations. To overcome these issues, a non standard generalization of the Ramsey protocol has been considered here and demonstrated additional benefits, sixty-five years after the original scheme was proposed. The Hyper-Ramsey spectral resonance has been fully extended in this letter to include potential biases, higher-order light-shift corrections on detunings and various modifications of laser parameters exploring non linear frequency responses of quantum particles for ultra precise frequency measurement.

The application of the Generalized Hyper-Ramsey resonance should be able to improve frequency uncertainty measurements in next tests of fundamental physics based on atomic or molecular fountains \cite{Beausoleil:1986,Bethlem:2008,Tarbutt:2013}, in charged ions \cite{Safronova:2014,Yudin:2014}, for small changes in molecular vibrational frequencies based on clocks sensitive to potential variation in the electron-to-proton mass ratio \cite{Shelkovnikov:2008,Schiller:2014} and in searching for a weak parity violation in chiral molecules by laser spectroscopy \cite{Tokunaga:2013}.
Quantum phase-shift engineering should impact high resolution mass spectrometry based on the application of the Ramsey method to short-lived ions stored in Penning traps \cite{George:2007} and laser pulsed spectroscopy in cold molecule chemistry \cite{Meerakker:2012,Heiner:2007}.
Using the Generalized Hyper-Ramsey spectroscopy for Stark decelerated cold molecules may allow a significant improvement of the frequency measurement uncertainty, which can be important for the search of the time-variation of the fine structure constant \cite{Hudson:2006}, in measuring gravitationally induced quantum phase shifts for neutrons~\cite{Abele:2010} and for observing spin dependent nuclear scattering lengths of neutrons in Ramsey interferometers \cite{Piegsa:2008}.
In the future, a pair of stretched hyperfine states from a $^{229}$Th nuclear transition may provide a large suppression of several external field shifts \cite{Peik:2002,Campbell:2012} where an ultra-narrow clock transition will offer an exquisite test of nuclear quantum engineering spectroscopy at the next level of $10^{-19}$ relative accuracy.

\section*{Acknowldegments}

\indent T. Zanon-Willette deeply acknowledges C. Janssen for checking calculations, J. Ye, E. Arimondo, J. Dalibard, B. Darquié, M. Glass-Maujean, M. Minissale, R. Metzdorff, A.D. Ludlow, E. de Clercq, M-L. Dubernet-Tuckey and Y. Té for discussions and a careful reading of the manuscript.

V.I.Yu. and A.V.T. were supported by the RFBR (grants 14-02-00712, 14-02-00939, 14-02-00806), by the Russian Academy of Sciences, by Presidium of the Siberian Branch of the Russian Academy of Sciences, by the RF Ministry of Education and Science (state assignment No. 2014/139 project No. 825).

\section*{Appendix}

\subsection{Wave-function model including phase of the laser field}

We present the wave-function formalism \cite{Ramsey:1950,Zanon:2014,Rabi:1945} to establish Eqs.~\ref{Generalized-Hyper-Ramsey-transition} to \ref{Generalized-Hyper-Ramsey-phase} in the main section of the paper. It is based on a two-level system describing the superposition of the $|g\rangle,|e\rangle$ clock states as
\begin{equation}
|\Psi(\theta_{l})\rangle=c_g(\theta_{l})|g\rangle+c_e(\theta_{l})|e\rangle.
\end{equation}
where the pulse area is defined by $\theta_{l}=\omega_{l}\tau_{l}/2$ and the effective Rabi field is $\omega_{l}^{2}=\delta_{l}^{2}+\Omega_{l}^{2}$.
Using the solution of the Schrödinger's equation, we write for the $c_{g,e}(\theta_{l})$ ($l=i,j,k$) transition amplitudes
\begin{equation}
\begin{split}
\left(%
\begin{array}{c}
c_{g}(\theta_{l})
\\
c_{e}(\theta_{l})
\\
\end{array}%
\right)
=\chi(\theta_{l})\cdot M(\theta_{l})\cdot
\left(%
\begin{array}{c}
c_{g}(0) \\
c_{e}(0) \\
\end{array}%
\right)
\end{split}
\end{equation}
including a phase factor of the form $\chi(\theta_{l})=\exp\left[-i\delta_{l}\frac{\tau_{l}}{2}\right]$.
The wave-function evolution driven by the pulse area $\theta_{l}$ is determined by the following complex 2x2 spinor interaction matrix as \cite{Varshalovich:1988,Rabi:1954,Jaynes:1955,Goldstein:1980}:
\begin{equation}
\begin{split}
M(\theta_{l})&=\left(
\begin{array}{cc}
M_{+}(\theta_{l}) &e^{\imath\varphi_{l}}M_{\dagger}(\theta_{l}) \\
e^{-\imath\varphi_{l}}M_{\dagger}(\theta_{l}) & M_{-}(\theta_{l}) \\
\end{array}%
\right)\\
&=\left(%
\begin{array}{cc}
\cos(\theta_{l})+i\frac{\delta_{l}}{\omega_{l}}\sin(\theta_{l})&-ie^{i\varphi_{l}}\frac{\Omega_{l}}{\omega_{l}}\sin(\theta_{l}) \\
-ie^{-i\varphi_{l}}\frac{\Omega_{l}}{\omega_{l}}\sin(\theta_{l})&\cos(\theta_{l})-i\frac{\delta_{l}}{\omega_{l}}\sin(\theta_{l}) \\
\end{array}
\right).
\end{split}
\label{matrix}
\end{equation}
where the phase of the laser field is introduced by $\varphi_{l}$.
The pulsed excitation is written as a product of different matrices $M(\theta_{l})$ and a free evolution without laser light during a time T. Generalized Hyper-Ramsey expression for respective composite sequences $\theta_{i},\delta T,\theta_{j},\theta_{k}$ and $\theta_{i},\theta_{j},\delta T,\theta_{k}$ which depend on initials conditions $c_{g}(0)$ and $c_{e}(0)$ are:
\begin{widetext}
\begin{equation}
\begin{split}
&\left(%
\begin{array}{c}
c_{g}(\theta_{i},\delta T,\theta_{j},\theta_{k}) \\
c_{e}(\theta_{i},\delta T,\theta_{j},\theta_{k}) \\
\end{array}%
\right)=\chi(\theta_{i},\theta_{j},\theta_{k})\cdot\left(%
\begin{array}{cc}
\begin{split}& M_{+}(\theta_{i})M_{+}(\theta_{j},\theta_{k})\\+&M_{\dagger}(\theta_{j},\theta_{k})M_{\dagger}(\theta_{i})e^{\left[-i(\delta T+\varphi_{i})\right]}\end{split}& \begin{split}&M_{\dagger}(\theta_{i})e^{i\varphi_{i}}M_{+}(\theta_{j},\theta_{k})\\+&M_{\dagger}(\theta_{j},\theta_{k})M_{-}(\theta_{i})e^{\left[-i\delta T\right]}\end{split}
\\\\
\begin{split}& M_{\dagger}(\theta_{k},\theta_{j})M_{+}(\theta_{i})\\+&M_{\dagger}(\theta_{i})M_{-}(\theta_{j},\theta_{k})e^{\left[-i(\delta T+\varphi_{i})\right]}\end{split}& \begin{split}&M_{\dagger}(\theta_{k},\theta_{j})M_{\dagger}(\theta_{i})e^{i\varphi_{i}}\\+&M_{-}(\theta_{j},\theta_{k})M_{-}(\theta_{i})e^{\left[-i\delta T\right]}\end{split} \\
\end{array}%
\right)\cdot\left(%
\begin{array}{c}
c_{g}(0) \\
c_{e}(0) \\
\end{array}%
\right)\label{sequence-a}\\
\end{split}
\end{equation}
\begin{equation}
\begin{split}
&\left(%
\begin{array}{c}
c_{g}(\theta_{i},\theta_{j},\delta T,\theta_{k}) \\
c_{e}(\theta_{i},\theta_{j},\delta T,\theta_{k}) \\
\end{array}%
\right)=\chi(\theta_{i},\theta_{j},\theta_{k})\cdot\left(%
\begin{array}{cc}
\begin{split}& M_{+}(\theta_{k})M_{+}(\theta_{i},\theta_{j})\\+&M_{\dagger}(\theta_{i},\theta_{j})M_{\dagger}(\theta_{k})e^{\left[-i(\delta T-\varphi_{k})\right]}\end{split}& \begin{split}&M_{\dagger}(\theta_{j},\theta_{i})M_{+}(\theta_{k})\\+&M_{\dagger}(\theta_{k})M_{-}(\theta_{i},\theta_{j})e^{\left[-i(\delta T-\varphi_{k})\right]}\end{split}
\\\\
\begin{split}& M_{\dagger}(\theta_{k})e^{-i\varphi_{k}}M_{+}(\theta_{i},\theta_{j})\\+&M_{\dagger}(\theta_{i},\theta_{j})M_{-}(\theta_{k})e^{\left[-i\delta T\right]}\end{split}& \begin{split}&M_{\dagger}(\theta_{j},\theta_{i})M_{\dagger}(\theta_{k})e^{-i\varphi_{k}}\\+&M_{-}(\theta_{i},\theta_{j})M_{-}(\theta_{k})e^{\left[-i\delta T\right]}\end{split} \\
\end{array}%
\right)\cdot\left(%
\begin{array}{c}
c_{g}(0) \\
c_{e}(0) \\
\end{array}%
\right)\label{sequence-b},
\end{split}
\end{equation}
\end{widetext}
with $\chi(\theta_{i},\theta_{j},\theta_{k})=\chi(\theta_{i})\chi(\theta_{j})\chi(\theta_{k})$ and where reduced matrix components are
\begin{equation}
\begin{split}
M_{+}(\theta_{j},\theta_{k})&=M_{+}(\theta_{j})M_{+}(\theta_{k})+M_{\dagger}(\theta_{j})M_{\dagger}(\theta_{k})e^{-i(\varphi_{j}-\varphi_{k})}, \\
M_{-}(\theta_{j},\theta_{k})&=M_{-}(\theta_{j})M_{-}(\theta_{k})+M_{\dagger}(\theta_{j})M_{\dagger}(\theta_{k})e^{i(\varphi_{j}-\varphi_{k})},\\
M_{\dagger}(\theta_{j},\theta_{k})&= M_{\dagger}(\theta_{j})e^{i\varphi_{j}}M_{+}(\theta_{k})+M_{\dagger}(\theta_{k})e^{i\varphi_{k}}M_{-}(\theta_{j}),\\
M_{\dagger}(\theta_{k},\theta_{j})&= M_{\dagger}(\theta_{k})e^{-i\varphi_{k}}M_{+}(\theta_{j})+M_{\dagger}(\theta_{j})e^{-i\varphi_{j}}M_{-}(\theta_{k}),
\end{split}
\label{reduced-components-a}
\end{equation}
\begin{equation}
\begin{split}
M_{+}(\theta_{i},\theta_{j})&=M_{+}(\theta_{i})M_{+}(\theta_{j})+M_{\dagger}(\theta_{i})M_{\dagger}(\theta_{j})e^{-i(\varphi_{i}-\varphi_{j})},\\
M_{-}(\theta_{i},\theta_{j})&=M_{-}(\theta_{i})M_{-}(\theta_{j})+M_{\dagger}(\theta_{i})M_{\dagger}(\theta_{j})e^{i(\varphi_{i}-\varphi_{j})},\\
M_{\dagger}(\theta_{i},\theta_{j})&= M_{\dagger}(\theta_{j})e^{-i\varphi_{j}}M_{+}(\theta_{i})+M_{\dagger}(\theta_{i})e^{-i\varphi_{i}}M_{-}(\theta_{j}),\\
M_{\dagger}(\theta_{j},\theta_{i})&= M_{\dagger}(\theta_{i})e^{i\varphi_{i}}M_{+}(\theta_{j})+M_{\dagger}(\theta_{j})e^{i\varphi_{j}}M_{-}(\theta_{i}).
\end{split}
\label{reduced-components-b}
\end{equation}
The final expression is an Hyper-Ramsey complex amplitude.
We are now able to explicit the transition probability $P_{|g\rangle\mapsto|e\rangle}$ for the two composite sequences.\\
For the first composite sequence $\theta_{i},\delta T,\theta_{j},\theta_{k}$ shown in Fig.~\ref{sequence}(a), we have:
\begin{subequations}
\begin{align}
P_{|g\rangle\mapsto|e\rangle}&=c_{e}(\theta_{i},T,\theta_{j},\theta_{k})c_{e}^{*}(\theta_{i},T,\theta_{j},\theta_{k}),\\
&=|\alpha|^{2}\left|1+\beta e^{-i(\delta T-\Phi+\varphi_{i})}\right|^{2}\label{probability-sequence-a}.
\end{align}
\end{subequations}
where wave-function envelops $\alpha,\beta$ driving the resonance amplitude are:
\begin{subequations}
\begin{align}
\alpha=&\left[M_{+}(\theta_{i})c_{g}(0)+M_{\dagger}(\theta_{i})e^{i\varphi_{i}}c_{e}(0)\right]M_{\dagger}(\theta_{k},\theta_{j})\nonumber\\
&\times\chi(\theta_{i},\theta_{j},\theta_{k})\label{envelop-sequence-a},\\
\beta e^{i\Phi}=&\left[\frac{M_{\dagger}(\theta_{i})c_{g}(0)+M_{-}(\theta_{i})c_{e}(0)}{M_{+}(\theta_{i})c_{g}(0)+M_{\dagger}(\theta_{i})e^{i\varphi_{i}}c_{e}(0)}\right]
\frac{M_{-}(\theta_{j},\theta_{k})}{M_{\dagger}(\theta_{k},\theta_{j})}\label{envelop-sequence-a}.
\end{align}
\end{subequations}\\
For the second composite sequence $\theta_{i},\theta_{j},\delta T,\theta_{k}$  shown in Fig.~\ref{sequence}(b), we also have:
\begin{subequations}
\begin{align}
P_{|g\rangle\mapsto|e\rangle}&=c_{e}(\theta_{i},\theta_{j},T,\theta_{k})c_{e}^{*}(\theta_{i},\theta_{j},T,\theta_{k}),\\
&=|\alpha|^{2}\left|1+\beta e^{-i(\delta T-\Phi-\varphi_{k})}\right|^{2}\label{probability-sequence-b}.
\end{align}
\end{subequations}
where envelops $\alpha,\beta$ driving the resonance amplitude are now:
\begin{subequations}
\begin{align}
\alpha=&\left[M_{+}(\theta_{i},\theta_{j})c_{g}(0)+M_{\dagger}(\theta_{j},\theta_{i})c_{e}(0)\right]M_{\dagger}(\theta_{k})e^{-i\varphi_{k}}\nonumber\\
&\times\chi(\theta_{i},\theta_{j},\theta_{k})\label{envelop-sequence-b},\\
\beta e^{i\Phi}=&\left[\frac{M_{\dagger}(\theta_{i},\theta_{j})c_{g}(0)+M_{-}(\theta_{i},\theta_{j})c_{e}(0)}{M_{+}(\theta_{i},\theta_{j})c_{g}(0)+M_{\dagger}(\theta_{j},\theta_{i})c_{e}(0)}\right]
\frac{M_{-}(\theta_{k})}{M_{\dagger}(\theta_{k})}\label{envelop-sequence-b}.
\end{align}
\end{subequations}
In all cases, the phase term $\Phi$ represents the atomic phase-shift accumulated by the wave-function during the laser interrogation sequence.
Starting from an initial condition $c_{g}(0)=1$ and $c_{e}(0)=0$, phase-shift expressions for sequences presented in Fig.~\ref{sequence}(a) and (b) are respectively given by:
\begin{subequations}
\begin{align}
\Phi&=Arg\left[\frac{M_{\dagger}(\theta_{i})}{M_{+}(\theta_{i})}\frac{M_{-}(\theta_{j},\theta_{k})}
{M_{\dagger}(\theta_{k},\theta_{j})}\right]\label{shift-a},\\
\Phi&=Arg\left[\frac{M_{\dagger}(\theta_{i},\theta_{j})}{M_{+}(\theta_{i},\theta_{j})}\frac{M_{-}(\theta_{k})}{M_{\dagger}(\theta_{k})}\right]\label{shift-b}.
\end{align}
\end{subequations}
When phases of laser fields are ignored in Eq.~\ref{reduced-components-a} and Eq.~\ref{reduced-components-b}, these expressions lead to the analytical form given by Eq.~\ref{Generalized-Hyper-Ramsey-phase}. It is determining the clock frequency shift measured on the central fringe in a Generalized Hyper-Ramsey spectroscopy.
Note that by applying a specific phase modulation of the laser field with $\varphi_{i}=\pm\pi/2,\varphi_{j}=\pi,\varphi_{k}=0$ ($\varphi_{i}=0,\varphi_{j}=\pi,\varphi_{k}=\mp\pi/2$) in Eq.~\ref{sequence-a}, Eq.~\ref{reduced-components-a}, Eq.~\ref{probability-sequence-a} (Eq.~\ref{sequence-b}, Eq.~\ref{reduced-components-b}, Eq.~\ref{probability-sequence-b}) then subtracting both components with opposite sign, we obtain the dispersive curve reported in Fig.~\ref{shift-comparison}(a).

\subsection{Ramsey transition probability ($\theta,\delta T,\theta$)}

We derive the analytical expression for the standard Ramsey transition probability from the generalized expression established in the main section.
We plug into in Eqs.~\ref{Generalized-Hyper-Ramsey-transition} to \ref{Generalized-Hyper-Ramsey-phase} the values for the Ramsey case from Table 1 with $\theta_j=0$ and $\theta_i=\theta_k=\theta$ where Eq.~\ref{reduced-beta} is reduced to one and Eq.~\ref{reduced-variable} is zero.
The generalized transition probability from state $|g\rangle$ to state $|e\rangle$ takes the following form:
\begin{equation}
\begin{split}
P_{|g\rangle\mapsto|e\rangle}=&2\frac{\Omega^{2}}{\omega^{2}}\sin^{2}\theta\left(\cos^{2}\theta+\frac{\delta^{2}}{\omega^{2}}\sin^{2}\theta\right)\\
&\times\left[1+\cos\left(\delta T+\Phi\right)\right],
\end{split}
\label{Ramsey-transition}
\end{equation}
where the Ramsey phase-shift is found to be \cite{Abramowitz:1968}:
\begin{equation}
\begin{split}
\Phi&=\arctan\left[\frac{2\frac{\delta}{\omega}\tan\theta}{1-\left(\frac{\delta}{\omega}\right)^2\tan^2\theta}\right]\\
&=2\arctan\left[\frac{\delta}{\omega}\tan\theta\right].
\end{split}
\label{Ramsey-shift}
\end{equation}
By applying a trigonometrical transformation on the following part
\begin{equation}
\begin{split}
&\left(\cos^{2}\theta+\frac{\delta^{2}}{\omega^{2}}\sin^{2}\theta\right)\times\left[1+\cos\left(\delta T+\Phi\right)\right]\\
&=2\left[\frac{\cos\theta+i\frac{\delta}{\omega}\sin\theta}{2}e^{i\frac{\delta T}{2}}+\frac{\cos\theta-i\frac{\delta}{\omega}\sin\theta}{2}e^{-i\frac{\delta T}{2}}\right]^2\\
&=2\left[\cos\left(\frac{\delta T}{2}\right)\cos\theta-\frac{\delta}{\omega}\sin\left(\frac{\delta T}{2}\right)\sin\theta\right]^2,
\end{split}
\end{equation}
we recover the standard expression of the transition probability derived by Ramsey in 1950 for a spin $1/2$ interacting with a radio-frequency field \cite{Ramsey:1950,Ramsey:1956} as:
\begin{equation}
\begin{split}
P_{|g\rangle\mapsto|e\rangle}=4\frac{\Omega^{2}}{\omega^{2}}\sin^{2}\theta&\left[\cos\left(\frac{\delta T}{2}\right)\cos\theta\right.\\
&\left.-\frac{\delta}{\omega}\sin\left(\frac{\delta T}{2}\right)\sin\theta\right]^{2}.
\end{split}
\label{Ramsey-1950}
\end{equation}
where $\theta=\omega\tau/2$.


\begin{thebibliography}{1}

\bibitem{Rosenband:2008} T. Rosenband, D.B. Hume, P.O. Schmidt, C.W. Chou, A. Brusch, L. Lorini, W.H. Oskay, R.E. Drullinger, T.M. Fortier, J.E. Stalnaker, S.A. Diddams, W.C. Swann, N.R. Newbury, W.M. Itano, D.J. Wineland, J.C. Bergquist, Science \textbf{319}, 1808 (2008).
\bibitem{Wineland:2013} D.J. Wineland, Rev. Mod. Phys. \textbf{85}, 1103 (2013).
\bibitem{Hinkley:2013} N. Hinkley, J.A. Sherman, N.B. Phillips, M. Schioppo, N.D. Lemke, K. Beloy, M. Pizzocaro, C.W. Oates and A.D. Ludlow, Science \textbf{341}, 1215 (2013).
\bibitem{Bloom:2014} B.J. Bloom, T.L. Nicholson, J.R. Williams, S.L. Campbell, M. Bishof, X. Zhang, W. Zhang, S.L. Bromley and J. Ye, Nature \textbf{506}, 71 (2014).
\bibitem{LeTargat:2013} R. Le Targat, L. Lorini, Y. Le Coq, M. Zawada, J. Guéna, M. Abgrall, M. Gurov, P. Rosenbusch, D.G. Rovera, B. Nagórny,	R. Gartman,	P.G. Westergaard, M.E. Tobar, M. Lours,	 G. Santarelli, A. Clairon, S. Bize, P. Laurent, P. Lemonde and J. Lodewyck, Nature Comm. \textbf{4}, 2109 (2013).
\bibitem{Brune:1990} M. Brune, S. Haroche, V. Lefevre, J. M. Raimond and N. Zagury, Phys. Rev. Lett. \textbf{65}, 976 (1990).
\bibitem{Gleyzes:2007} S. Gleyzes, S. Kuhr, C. Guerlin, J. Bernu, S. Deléglise, U. Busk Hoff, M. Brune, J.-M. Raimond and Serge Haroche, Nature \textbf{446}, 297 (2007).
\bibitem{Haroche:2013} S. Haroche, Rev. Mod. Phys. \textbf{85}, 1083 (2013).
\bibitem{Cronin:2005} A.D. Cronin, J. Schmiedmayer, and D.E. Pritchard, Rev. Mod. Phys. \textbf{81}, 1051 (2005).
\bibitem{Berman:1996} P.R. Berman, \textit{Atom Interferometry}, Academic Press, (1996).
\bibitem{Klepp:2014} J. Klepp, S. Sponar, and Y. Hasegawa, Prog. Theor. Exp. Phys, 082A01 (2014).
\bibitem{Rabi:1938} I. I. Rabi, J. R. Zacharias, S. Millman and P. Kusch, Phys. Rev. \textbf{53}, 318 (1938).
\bibitem{Ramsey:1950} N.F. Ramsey, Phys. Rev. \textbf{78}, 695 (1950).
\bibitem{Ramsey:1956} N.F. Ramsey, \textit{Molecular beams}, Clarendon Press, Oxford (1956).
\bibitem{Ramsey:1990} N.F. Ramsey, Rev. Mod. Phys. \textbf{62}, 541 (1990).
\bibitem{Vanier:1989} J. Vanier and C. Audoin, \textit{The quantum physics of atomic frequency standards}, Adam Hilger IOP, Bristol, (1989).
\bibitem{Gibble:1992} K. Gibble and S. Chu, Metrologia \textbf{29}, 201 (1992).
\bibitem{Wynands:2005} R. Wynands and S. Weyers, Metrologia \textbf{42}, S64 (2005).
\bibitem{Bize:2005} S. Bize, P. Laurent, M. Abgrall, H. Marion, I. Maksimovic, L. Cacciapuoti, J. Grünert, C. Vian, F. Pereira dos Santos, P. Rosenbusch, P. Lemonde, G. Santarelli, P. Wolf, A. Clairon, A. Luiten, M. Tobar and C. Salomon, J. Phys. B \textbf{38}, S449 (2005).
\bibitem{Margolis:2009} H.S. Margolis, Eur. Phys. J. Special Topics \textbf{172}, 97 (2009).
\bibitem{DereviankoKatori:2011} A. Derevianko and H. Katori, Rev. Mod. Phys. \textbf{83}, 331 (2011).
\bibitem{Ludlow:2015} A.D. Ludlow, M.M. Boyd, J. Ye, E. Peik and P.O. Schmidt, Rev. Mod. Phys. \textbf{87}, 637 (2015).
\bibitem{Nicholson:2015} T.L. Nicholson, S.L. Campbell, R.B. Hutson, G.E. Marti, B.J. Bloom, R.L. McNally, W. Zhang, M.D. Barrett, M.S. Safronova, G.F. Strouse, W.L. Tew and J. Ye, Nature Comm. \textbf{6}, 7896 (2015).
\bibitem{Campbell:2009} G.K. Campbell, M.M. Boyd, J.W. Thomsen, M.J. Martin, S. Blatt, M.D. Swallows, T.L. Nicholson, T. Fortier, C.W. Oates, S.A. Diddams, N.D. Lemke, P. Naidon, P. Julienne, Jun Ye and A.D. Ludlow, Science \textbf{324}, 360 (2009).
\bibitem{Lisdat:2009} Ch. Lisdat, J.S.R. Vellore Winfred, T. Middelmann, F. Riehle and U. Sterr, Phys. Rev. Lett. \textbf{103}, 090801 (2009).
\bibitem{Chou:2010} C.W. Chou, D.B. Hume, T. Rosenband and D.J. Wineland, Science \textbf{329}, 1630 (2010).
\bibitem{Blaum:2006} K. Blaum, Phys. Rep. \textbf{425}, 1 (2006).
\bibitem{Bollen:1992} G. Bollen, H.-J. Kluge, T. Otto, G. Savard and H. Stolzenberg, Nucl. Instrum. Meth. \textbf{B70}, 490 (1992).
\bibitem{George:2007} S. George, S. Baruah, B. Blank, K. Blaum, M. Breitenfeldt, U. Hager, F. Herfurth, A. Herlert, A. Kellerbauer, H.-J. Kluge, M. Kretzschmar, D. Lunney, R. Savreux, S. Schwarz, L. Schweikhard and C. Yazidjian, Phys. Rev. Lett. \textbf{98}, 162501 (2007).
\bibitem{Eibach:2011} M. Eibach, T. Beyer, K. Blaum, M. Block, K. Eberhardt, F. Herfurth, J. Ketelaer, Sz. Nagy, D. Neidherr, W. Nörtershäuser and C. Smorra, Int. J. Mass Spectrom. \textbf{303} 27 (2011).
\bibitem{Yudin:2010} V.I. Yudin, A.V. Taichenachev, C.W. Oates, Z.W. Barber, N.D. Lemke, A.D. Ludlow, U. Sterr, Ch. Lisdat and F. Riehle, Phys. Rev. A \textbf{82}, 011804(R) (2010).
\bibitem{Tabatchikova:2013} K.S.Tabatchikova, A. V. Taichenachev, V. I. Yudin, JETP Lett. \textbf{97}, 311 (2013).
\bibitem{Tabatchikova:2015} K.S.Tabatchikova, A. V. Taichenachev, A. K. Dmitriev, V. I. Yudin, JETP Lett.
\textbf{120}, 203 (2015).
\bibitem{Levitt:1986} M.H. Levitt, \textit{Composite pulses}, Prog. Nucl. Mag. Res. Spect. \textbf{18}, 61 (1986).
\bibitem{Vandersypen:2005} L.M.K. Vandersypen and I.L. Chuang, Rev. Mod. Phys. \textbf{76}, 1037 (2005).
\bibitem{Braun:2014} M. Braun and S.J. Glaser, New J. Phys. \textbf{16}, 115002 (2014).
\bibitem{Hahn:1950} E.L. Hahn, Phys. Rev. \textbf{80}, 580 (1950).
\bibitem{Carr:1958} H.Y. Carr, Phys. Rev. \textbf{112}, 1693 (1958).
\bibitem{Huntemann:2012} N. Huntemann, B. Lipphardt, M. Okhapkin, Chr. Tamm, E. Peik, A.V. Taichenachev and V.I. Yudin, Phys. Rev. Lett. \textbf{109}, 213002 (2012).
\bibitem{Zanon:2014} T. Zanon-Willette, S. Almonacil, E. de Clercq, A.D. Ludlow and E. Arimondo, Phys. Rev. A \textbf{90}, 053427 (2014).
\bibitem{Rabi:1945} F. Bloch and I.I. Rabi, Rev. Mod. Phys. \textbf{17}, 237 (1945).
\bibitem{Varshalovich:1988} D.A Varshalovich, A.N. Moskalev and V.K. Khersonskii, \textit{Quantum Theory of Angular Momentum}, Singapore: World Scientific, (1988).
\bibitem{Kasevich:1991} M. Kasevich and S. Chu, Phys. Rev. Lett. \textbf{67}, 181 (1991).
\bibitem{Moler:1992} K. Moler, D.S. Weiss, M. Kasevich, and S. Chu, Phys. Rev. A \textbf{45}, 342 (1992).
\bibitem{Taichenachev:2006} A.V. Taichenachev, V.I. Yudin, C.W. Oates, C.W. Hoyt, Z.W. Barber and L. Hollberg, Phys. Rev. Lett. \textbf{96}, 083001 (2006), Z. Barber, C. Hoyt, C. Oates, L. Hollberg, A. Taichenachev and V. Yudin, Phys. Rev. Lett. \textbf{96}, 083002 (2006).
\bibitem{Kretzschmar:2007} M. Kretzschmar, Int. J. Mass Spectrom. \textbf{264}, 122 (2007).
\bibitem{Shirley:1963} J.H. Shirley, J. Appl. Phys. \textbf{34}, 783 (1963).
\bibitem{Greene:1978} G.L. Greene, Phys. Rev. A \textbf{18}, 1057 (1978).
\bibitem{Brandin:1994} A.B. Brandin, Phys. Rev. A \textbf{50}, 1575 (1994).
\bibitem{Rabi:1954} I.I. Rabi, N.F. Ramsey, J. Schwinger, Rev. Mod. Phys. \textbf{26}, 167 (1954).
\bibitem{Jaynes:1955} E.T. Jaynes, Phys. Rev. \textbf{98}, 1099 (1955).
\bibitem{Goldstein:1980} H. Goldstein, \textit{Classical Mechanics}, 2nd ed. Reading, MA: Addison-Wesley, (1980).
\bibitem{Feynman:1957} R.P. Feynman, F.L. Vernon, and R.W. Hellwarth, J. Appl. Phys. \textbf{28}, 49 (1957).
\bibitem{Glass-Maujean:1991} M. Glass-Maujean, H. Henry Stroke, Am. J. Phys. \textbf{59}, 886 (1991).
\bibitem{Abramowitz:1968} M. Abramowitz and I.A. Stegun, {\it Handbook of mathematical functions} (Dover Publications, Inc., New York, 1968).
\bibitem{Ramsey:1951} Norman F. Ramsey and Henry B. Silsbee, Phys. Rev. \textbf{84}, 506 (1951).
\bibitem{Morinaga:1989} A. Morinaga, F. Riehle, J. Ishikawa and J. Helmcke, Appl. Phys. B \textbf{48}, 165 (1989), IEEE Trans. instrum. Meas. \textbf{38}, 524 (1989).
\bibitem{Letchumanan:2004} V. Letchumanan, P. Gill, E. Riis, and A. G. Sinclair,
Phys. Rev. A \textbf{70}, 033419 (2004).
\bibitem{Beausoleil:1986} R.G. Beausoleil and T.W. Hänsch, Phys. Rev. A \textbf{33}, 1661 (1986).
\bibitem{Bethlem:2008} H.L. Bethlem , M. Kajita, B. Sartakov, G. Meijer and W. Ubachs, Eur. Phys. J. Special Topics \textbf{163}, 55 (2008).
\bibitem{Tarbutt:2013} M.R. Tarbutt, B.E. Sauer, J.J. Hudson and E.A. Hinds, New J. Phys. \textbf{15}, 053034 (2013).
\bibitem{Safronova:2014} M.S. Safronova, V.A. Dzuba, V.V. Flambaum, U.I. Safronova, S.G. Porsev and M.G. Kozlov, Phys. Rev. Lett. \textbf{113}, 030801 (2014).
\bibitem{Yudin:2014} V.I. Yudin, A.V. Taichenachev and A. Derevianko, Phys. Rev. Lett. \textbf{113}, 233003 (2014).
\bibitem{Shelkovnikov:2008} A. Shelkovnikov, R.J. Butcher, C. Chardonnet and A. Amy-Klein, Phys. Rev. Lett. \textbf{100}, 150801 (2008).
\bibitem{Schiller:2014} S. Schiller, D. Bakalov and V.I. Korobov, Phys. Rev. Lett. \textbf{113}, 023004 (2014).
\bibitem{Tokunaga:2013} S. Tokunaga, S.C. Stoeffler, F. Auguste, A. Shelkovnikov, C. Daussy, A. Amy-Klein, C. Chardonnet and B. Darquié, Mol. Phys. \textbf{111} 2363 (2013).
\bibitem{Meerakker:2012} S.Y.T. van de Meerakker, H.L. Bethlem , N. Vanhaecke and G. Meijer, Chem. Rev. \textbf{112}, 4828 (2012).
\bibitem{Heiner:2007} C.E. Heiner, D. Carty, G. Meijer and H.L. Bethlem, Nat. Phys. \textbf{3}, 115 (2007).
\bibitem{Hudson:2006} E.R. Hudson, H.J. Lewandowski, B.C. Sawyer and J. Ye, Phys. Rev. Lett. \textbf{96}, 143004 (2006).
\bibitem{Abele:2010} H. Abele, T. Jenke, H. Leeb, and J. Schmiedmayer, Phys. Rev. D. \textbf{81}, 065019 (2010).
\bibitem{Piegsa:2008} F.M. Piegsa, B. van den Brandta, H. Glättlic, P. Hautlea, J. Kohlbrechera, J.A. Kontera, B.S. Schlimmed, O. Zimmerb, Nucl. Instrum. Meth. A, \textbf{589}, 318 (2008).
\bibitem{Peik:2002} E. Peik, C. Tamm, EuroPhys. Lett. \textbf{61}, 181 (2003).
\bibitem{Campbell:2012} C.J. Campbell, A.G. Radnaev, A. Kuzmich,  V.A. Dzuba, V.V. Flambaum and A. Derevianko, Phys. Rev. Lett. \textbf{108}, 120802 (2012).
\end{thebibliography}
\end{document}